%% file: main.tex
\documentclass[conference]{IEEEtran}
\IEEEoverridecommandlockouts
\usepackage{cite}
\usepackage{amsmath,amssymb,amsfonts}
\usepackage{algorithmic}
\usepackage{graphicx}
\usepackage{textcomp}
\usepackage{xcolor}
\usepackage{siunitx}
\usepackage{booktabs}
\usepackage{makecell}
\def\BibTeX{{\rm B\kern-.05em{\sc i\kern-.025em b}\kern-.08em
    T\kern-.1667em\lower.7ex\hbox{E}\kern-.125emX}}

\begin{document}

\sisetup{mode=match}

% \title{\textsc{Megatron}: A 28nm \textbf{\color{red}XX}~GOPS/W \textbf{\color{red}YY}~Mparam/mm${}^2$ Mixed PCM-based Analog CiM/Digital RISC-V System-on-Chip\\for Edge GenAI}
\title{\textsc{Megatron}: a 28nm Analog PCM CiM/Digital System-on-Chip for Edge GenAI at 57.5~TOPS/W and 1.52~Mparam/mm${}^2$}

\author{
\IEEEauthorblockN{Alessandro~Nadalini\IEEEauthorrefmark{1}\IEEEauthorrefmark{2},
Angelo~Garofalo\IEEEauthorrefmark{1}\IEEEauthorrefmark{2},
Lorenzo~Greco\IEEEauthorrefmark{1}\IEEEauthorrefmark{2},
Andrea~Belano\IEEEauthorrefmark{2}\IEEEauthorrefmark{3},
Alessio~Antolini\IEEEauthorrefmark{2},\\
Francesco~Zavalloni\IEEEauthorrefmark{2},
Andrea~Lico\IEEEauthorrefmark{3},
Riccardo~Zurla\IEEEauthorrefmark{4},
Emanuela~Calvetti\IEEEauthorrefmark{4},
Luigi~Croce\IEEEauthorrefmark{4},\\
Marco~Pasotti\IEEEauthorrefmark{4},
Alessandro~Cabrini\IEEEauthorrefmark{5},
Eleonora~Franchi~Scarselli\IEEEauthorrefmark{2},
Davide~Rossi\IEEEauthorrefmark{2}\IEEEauthorrefmark{3},
Francesco~Conti\IEEEauthorrefmark{2}\IEEEauthorrefmark{3}\\
\IEEEauthorblockA{
    \IEEEauthorrefmark{2}University of Bologna, Italy,
    \IEEEauthorrefmark{3}Chips-IT, Italy,
    \IEEEauthorrefmark{4}STMicroelectronics, Italy,
    \IEEEauthorrefmark{5}University of Pavia, Italy,
    \IEEEauthorrefmark{1}\textit{Equal~contribution}
}}%
}%

\maketitle

\begin{abstract}
% Edge generative AI (GenAI) aims at moving computation from datacenters to the edge devices themselves to better support the requirements of wearable and autonomous systems. 
We present \textsc{Megatron}, a heterogeneous Edge GenAI System-on-Chip in 28nm FD-SOI CMOS technology combining a non-volatile analog in-memory-computing engine based on a 4Mi-cell phase-change memory (PCM) array with a digital RISC-V-based flexible neural processing unit. \textsc{Megatron} demonstrates up to {3.5~TOPS/W} using the RISC-V processors and 57.5~TOPS/W with PCiM, at a storage density of 1.52~Mparam/mm${}^2$ with 4-bit effective weight precision.
\end{abstract}

% \begin{IEEEkeywords}
% Phase-Change Memory, Analog-in-Memory Computing, Edge AI, System-on-Chip
% \end{IEEEkeywords}

\newcommand{\matrixtau}{PCiM}

% INPUTS
\input{text/01_Introduction}

\input{text/02_SoC_architecture}

\input{text/03_Measurements_and_Benchmarking}

\input{text/04_conclusion}

\bibliographystyle{ieeetr}
\bibliography{text/bibliography}

\end{document}

%% file: text/01_Introduction.tex
\section{Introduction}
The recent, exponential growth in capabilities of generative artificial intelligence (GenAI) algorithms such as Large Language Models (LLMs) and Mamba/SSMs is bringing about a revolution in the capabilities of wearables, robots, and other autonomous devices.
Yet, the dependence on computation performed in remote datacenters raises many challenges in terms of latency, availability, and privacy.
Edge GenAI aims at moving at least part of the burden of execution from datacenters to the edge devices themselves, similarly to what was done in the past years with perception AI algorithms such as CNNs -- however, Edge GenAI algorithms, such as Small Language Models (SLMs), have much more challenging characteristics in terms of memory footprint (10-100$\times$ more parameters; need for space for key-value (KV) caching) and computational patterns (e.g., the decode phase of LLMs is typically memory-bound due to the combined effect of low weight reuse in matrix-vector products and KV caching).
% \begin{figure}[tb]
%     \centering
%     \includegraphics[width=0.95\linewidth]{images/placeholder.png}
%     \caption{Motivational figure}
%     \label{fig:placeholder0}
% \end{figure}

Non-volatile analog in-memory-computing (NV-AIMC) based on resistive memories (ReRAMs)~\cite{reram-transformer-isscc26} and phase-change memories (PCMs)~\cite{pcm-isscc22}~\cite{hermes-nature} is attractive for Edge GenAI: putting high density non-volatile parameter storage with computation effectively removes all memory traffic due to network weights, leaving only that of KV caching.
However, it also raises many challenges for system-level integration and scaling to advanced technologies.
Computational PCMs have recently attracted attention due to the possibility to achieve high weight density and circumvent data drift.
However, previous works have focused primarily on technological aspects while neglecting system-level impacts of the introduction of this technology.

To fill in this gap, we introduce \textsc{Megatron}, a prototype heterogeneous System-on-Chip in 28nm FD-SOI CMOS STMicroelectronics technology for Edge GenAI combining NV-AIMC based on a 4Mi-cells phase-change memories (PCMs) with a fully flexible and programmable digital neural processing unit (NPU) based on a RISC-V DSP cluster.
Our contributions are the following:
\begin{enumerate}
    \item we show industry-leading on-chip parameter density in 28nm node of 1.52 Mparam/mm${}^2$;
    \item we propose a mixed analog PCM/digital system, achieving up to {57.5 TOPS/W} on linear layers at system-level with 8-bit activations and ENOB=4 weights;
    \item we extrapolate end-to-end fully-on-chip execution of one layer of an emerging SLM (SmollLM-135M).
\end{enumerate}

%\nocite{DIANA}
%\nocite{falcon-esserc25}
%\nocite{desoli}
%\nocite{reram-on-logic-isscc26}
%\nocite{siracusa}

%% file: text/02_SoC_architecture.tex
\section{SoC Architecture}

\begin{figure}[tb]
    \centering
    \includegraphics[width=0.9\linewidth]{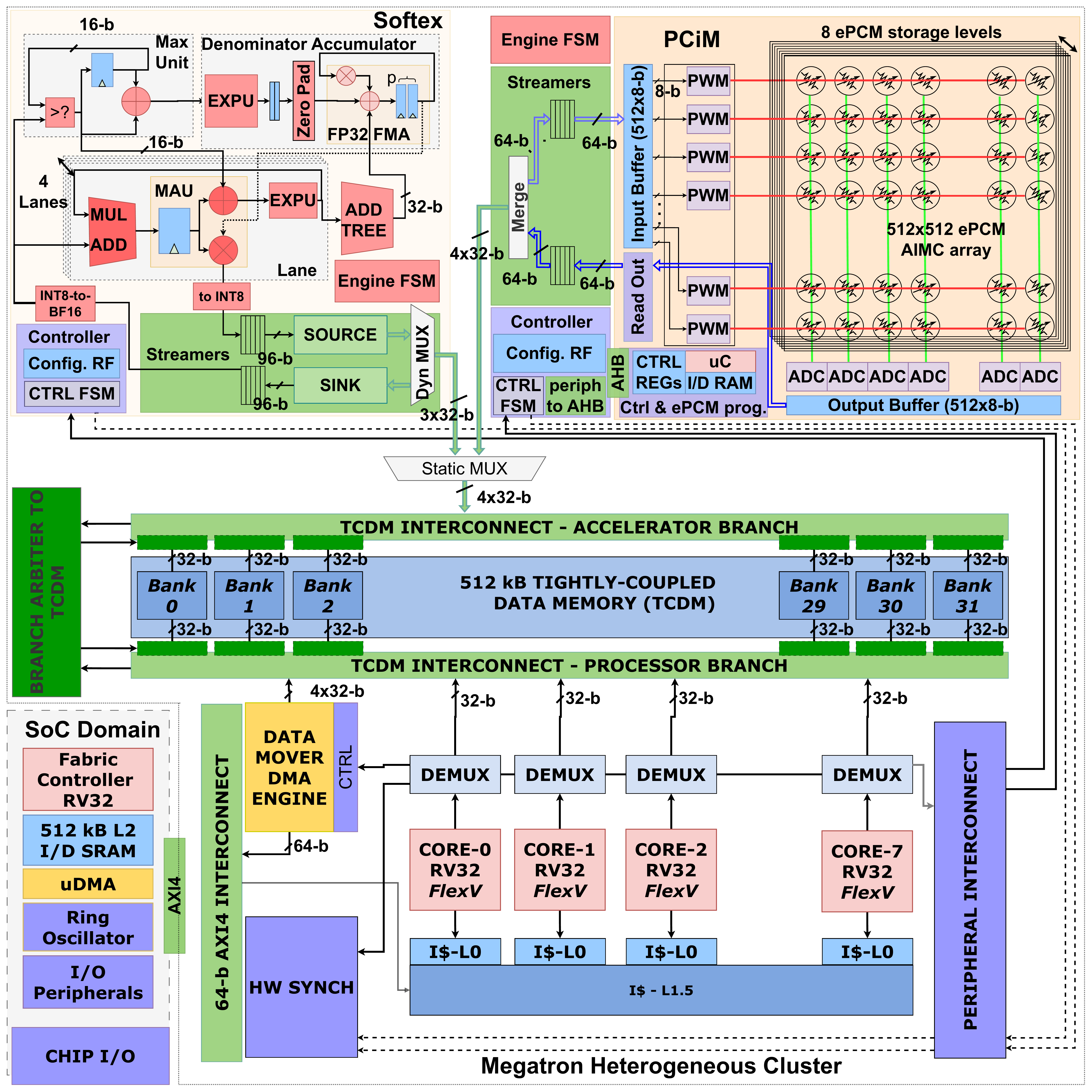}
    \caption{\textsc{Megatron} SoC architecture and Heterogeneous Cluster domain.}
    \label{fig:soc-archi}
\end{figure}

\textsc{Megatron} includes two domains: \textit{System-on-Chip (SoC)} and \textit{Heterogeneous Cluster (HC)}, as shown in Figure~\ref{fig:soc-archi}.
The SoC domain acts as an on-chip RISC-V \texttt{ RV32IMC} microcontroller for testing purposes, including I/O peripherals, an autonomous I/O uDMA, 512~KiB of L2 SRAM for data and instructions, and a ring oscillator for clock generation.
The HC domain comprises the primary contribution of our work, targeting on-device perceptive and generative Transformer models.
It includes eight RISC-V \texttt{ RV32IMCFXpulpNN} cores (\textbf{FlexV})~\cite{flex-v}, a DMA engine, a Phase-Change Compute-in-Memory accelerator (\textbf{\matrixtau{}}) and a digital accelerator for non-linear operators such as Softmax and GeLU (\textbf{Softex}).
All units are connected to a 32-bank, 512~KiB Tightly-Coupled Data Memory (TCDM) accessible through a low-latency (1-cycle) hybrid interconnect enabling simultaneous access from an accelerator branch (from \matrixtau{}, Softex) and a processor branch (from all FlexVs and the cluster DMA), forming a single tightly-coupled unit with a total of {512b/cycle} of aggregate bandwidth ({30.72~GB/s}~@~480~MHz).
Both Softex and \matrixtau{} access the TCDM accelerator branch through a set of \textit{streamers} that expose inputs and outputs to their respective microarchitecture as ready-valid data streams.
Both accelerators are controlled via a memory-mapped register interface; the access of Softex and \matrixtau{} is alternate and regulated by a static multiplexer.

The computational core of \matrixtau{} is a phase-change memory (PCM) analog in-memory computing (AIMC) macro \cite{Pasotti-esserc} in STMicroelectronics 28nm FDSOI CMOS, wrapped within a digital interface. It accelerates products between a stationary (non-volatile) matrix and a streamer-produced vector, exploiting analog storage and computation within the PCM array itself. \matrixtau{} stores 2M signed 4-bit weights (ENOB) across eight alternative \textit{scenarios}, supporting $512\times512$ matrix-vector products with signed 8-bit inputs and up to 11-bit output precision.
Each weight employs two PCM cells to encode positive and negative values.
During AIMC computation, inputs are encoded as time intervals on PCM Wordlines (WLs), while dedicated Bitline (BL) voltage regulators overcome PCM current-voltage non-linearity. BL currents are integrated over a time window and digitized via current-controlled oscillator ADCs. An internal 32-bit microcontroller handles configuration, weight programming, and calibration, while runtime execution uses a digital FSM interface. Double buffers on decoupled input/output channels enable pipelined dataflow with the streamers, supporting both GEMM and MVM kernels for the \textit{prefill} and \textit{decode} of autoregressive Transformers.

Softex is a hardware accelerator for non-linearities such as Softmax and GeLU, integrated as a cooperative accelerator sharing the TCDM with the FlexV cores and with PCiM. Its datapath features 4 lanes, each comprising a BF16 Multiplication-and-Addition Unit (MAU) and an Exponential Unit (EXPU) implementing a hardware-friendly approximation of the exponential function based on an extension of Schraudolph's method with polynomial mantissa correction. Softmax computation is organized in three phases: accumulation (max search and denominator calculation), inversion (Newton-Raphson reciprocal), and normalization (output probability computation). Softex communicates with the rest of the HC through a ready/valid streaming interface connected to the TCDM accelerator branch: configurable INT8-to-BF16 and BF16-to-INT8 data converters are used for compatibility with the 8-bit activation format of PCiM.

%% file: text/03_Measurements_and_Benchmarking.tex
\section{Measurements and Benchmarking}

\begin{figure}[tb]
    \centering
    \includegraphics[width=0.95\linewidth]{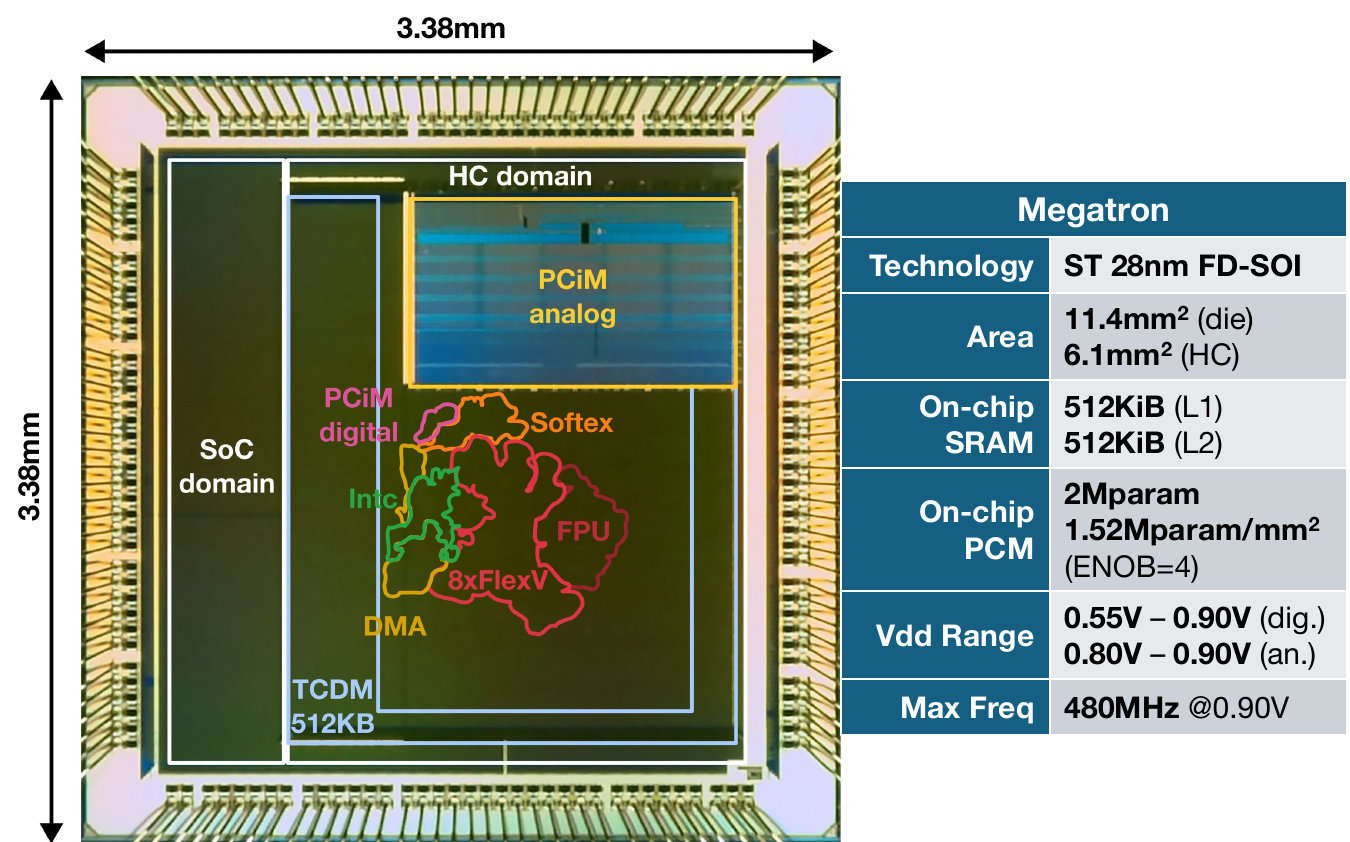}
    \caption{\textsc{Megatron} chip microphotograph, highlighting the SoC and HC domains and the main computing blocks within the HC.}
    \label{fig:chip_micrograph}
\end{figure}

Figure~\ref{fig:chip_micrograph} shows a microphotograph of the \textsc{Megatron} chip, fabricated in 28nm FD-SOI CMOS STMicroelectronics technology.
The die size is {3.38mm$\times$3.38mm}. We focused our analysis on the HC domain {(6.1mm${}^2$)}, which includes the \matrixtau{} macro ({1.34~mm${}^2$}, yielding a storage density of {1.49~Mparam/mm${}^2$}).
The same supply powers both the digital and analog IPs of \textsc{Megatron}.
The chip is working between \SI{0.55}{\volt}--\SI{0.90}{\volt} for what concerns the digital components and SRAM memories, and between \SI{0.80}{\volt}--\SI{0.90}{\volt} for the analog PCiM.
\begin{figure}[tb]
    \centering
    \includegraphics[width=0.95\linewidth]{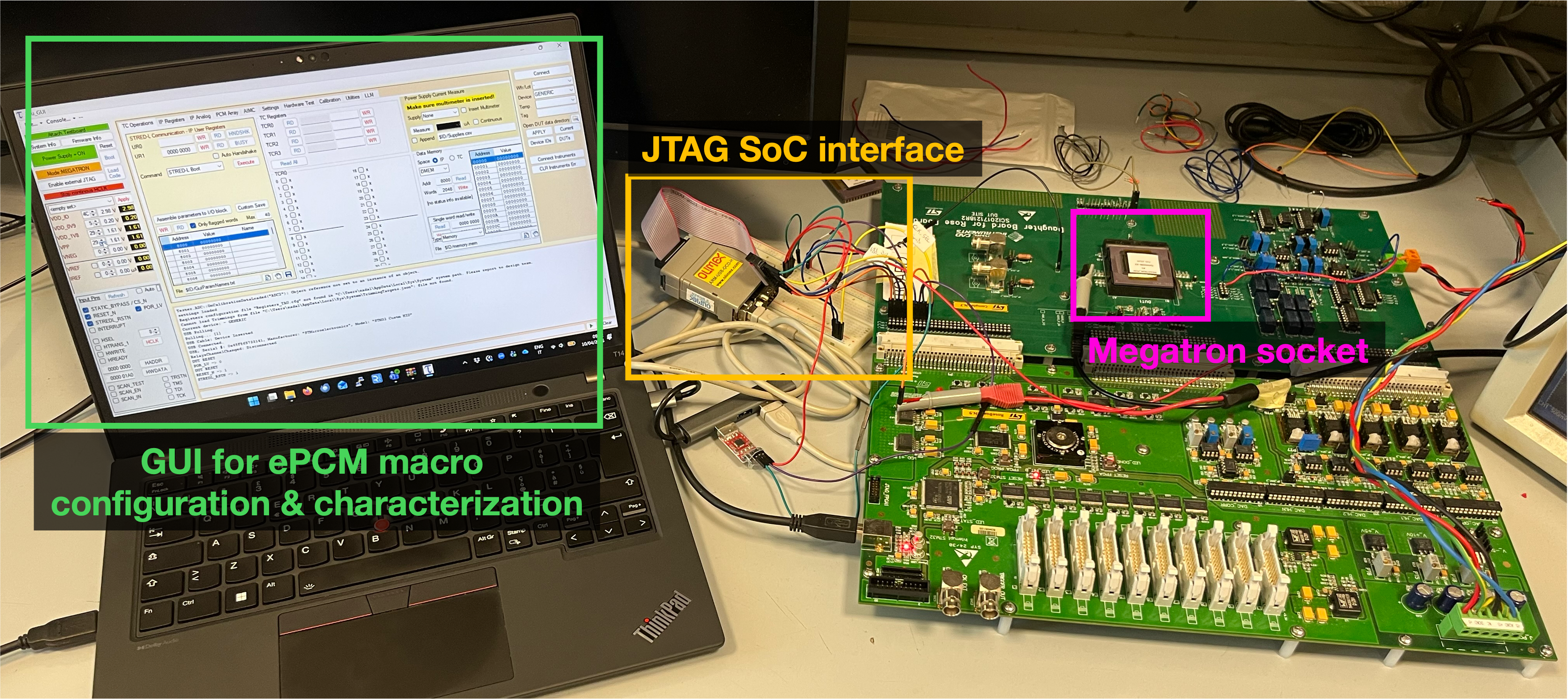}
    \caption{Experimental setup for \textsc{Megatron} test chip characterization.}
    \label{fig:experimental_setup}
\end{figure}
% 
% We characterized the digital part of the HC in terms of operating voltage, frequency, and power consumption at room temperature (\SI{25}{\celsius}) under a variety of workloads.
% Figure~\ref{fig:experimental_setup} shows our experimental setup: the \textsc{Megatron} prototype is mounted within a socket in a carrier board, designed specifically for the \textsc{Megatron} SoC, which exposes accessible pins to enable probing system I/Os with an oscilloscope and measuring current from the power supply with a digital multimeter.
% The carrier board is mounted on a mother board hosting an FPGA and other circuitry that enables characterization of the analog PCM macro from a laptop.
% The digital architecture of \textsc{Megatron} is also accessible by means of a JTAG interface connected to the SoC domain's debug module.
We characterized the HC digital part across operating voltages, frequencies, and workloads at room temperature (\SI{25}{\celsius}). As shown in Figure~\ref{fig:experimental_setup}, the \textsc{Megatron} prototype is socketed on a dedicated carrier board exposing probe points for oscilloscope measurements and current monitoring via digital multimeter. The carrier board mounts on a motherboard hosting an FPGA for PCM macro characterization from a laptop, with digital architecture access via a JTAG interface connected to the SoC debug module.

\begin{figure*}[tb]
    \centering
    \includegraphics[width=0.99\linewidth]{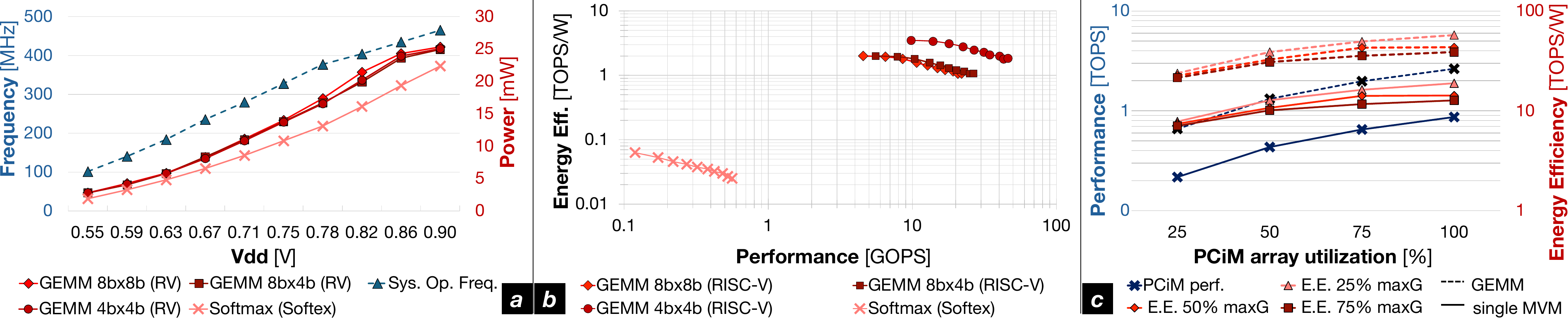}
    \caption{\textit{a)} Voltage/frequency/power sweep on \textsc{Megatron} prototype; \textit{b)} Performance and energy efficiency of digital blocks; \textit{c)} Performance and energy efficiency of PCiM sweeping conductance  (25--50--75\% of the maximum $G$), PCiM array utilization and kernel (GEMM vs single MVM). For the Softmax kernel we adopt the convention of 1 OP/token.}
    \label{fig:digital_sweep}
\end{figure*}

Figure~\ref{fig:digital_sweep}a reports experimental results in terms of frequency and HC domain power across the \SI{0.55}{\volt}--\SI{0.90}{\volt} supply voltage range, targeting a matrix multiplication kernel executed by the FlexV cores under several bit-width configurations and the operation of Softex.
\textsc{Megatron} can operate at a maximum frequency of \SI{480}{\mega\hertz}@\SI{0.90}{\volt} in the highest performance operating point, and at up to \SI{100}{\mega\hertz}@\SI{0.55}{\volt} in the slowest operating point.
% 
% \begin{figure}[tb]
%     \centering
%     \includegraphics[width=0.98\linewidth]{images/energy_efficiency.pdf}
%     \caption{Energy efficiency. For the Softmax kernel we adopt the convention of 1 OP/token.}
%     \label{fig:energy_efficiency}
% \end{figure}
% 
In Figure~\ref{fig:digital_sweep}b we characterize \textsc{Megatron} in terms of energy efficiency, considering the operation of digital IPs.
FlexV achieves up to 3.5 TOPS/W in the best-efficiency operating point (0.55V 4$\times$4-bit), providing a fully programmable NPU with efficiency comparable to custom datapath; in the best-performance point (0.90V), it achieves 46 GOPS at 1.84 TOPS/W.
Softex can process up to 560 Mtoken/s at 25 Gtoken/s/W at 0.90V, or up to 63 Gtoken/s/W at 0.55V.
Figure~\ref{fig:digital_sweep}c analyzes the performance and energy efficiency of the PCiM array while sweeping average conductance, PCM utilization and target kernel (GEMM vs MVM).
PCiM achieves up to 868~GOPS @ 18.9 TOPS/W in MVM and 2.64~TOPS @ 57.5 TOPS/W in GEMM.
% \textbf{\color{red} Briefly comment results here.}

\section{Evaluation on SmolLM-135M}
\begin{figure}[tb]
    \centering
    \includegraphics[width=0.98\linewidth]{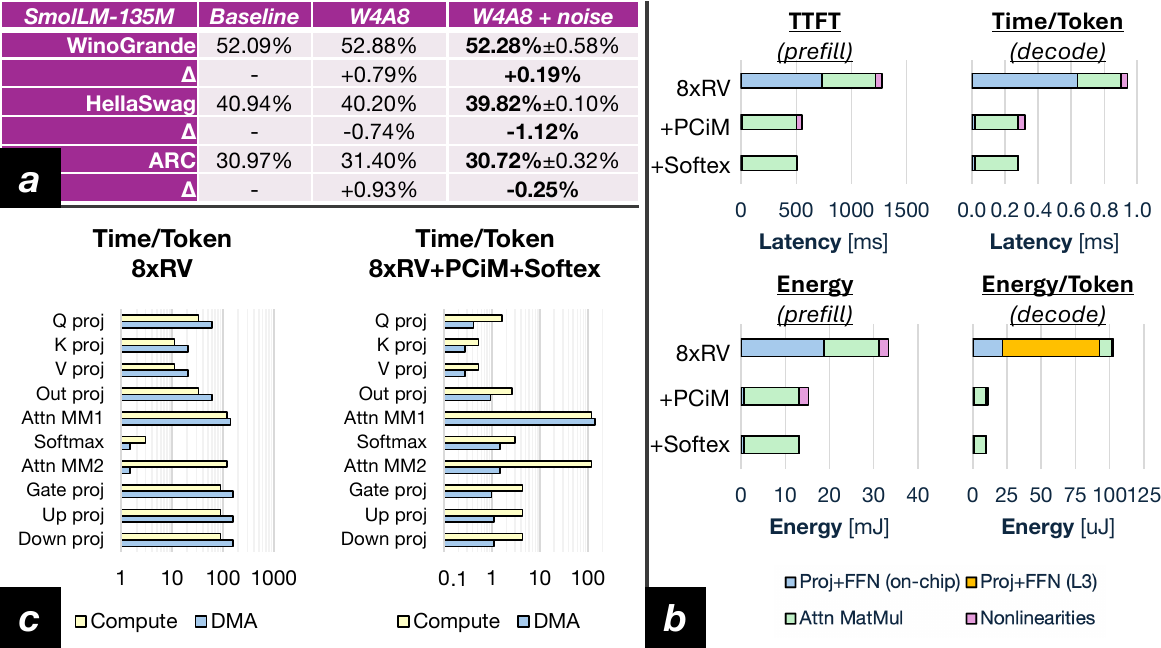}
    \caption{Evaluation of 1 layer of \textit{SmolLM-135M} with sequence size $S=2048$ on \textsc{Megatron-XL}: \textit{a)} accuracy on WinoGrande, HellaSwag and ARC-Challenge under full-precision, quantized (\textit{W4A8}), and analog (\textit{W4A8 + noise}) conditions; \textit{b)} latency and energy analysis in prefill and decode phase with/without on-chip PCiM and Softex; \textit{c)} detail of decode phase time-per-token for one SmolLM-135M layer.}
    \label{fig:end_to_end}
\end{figure}

% MOVED HERE FOR PAGE LAYOUT
\begin{figure*}[t]
    \centering
    \includegraphics[width=\linewidth]{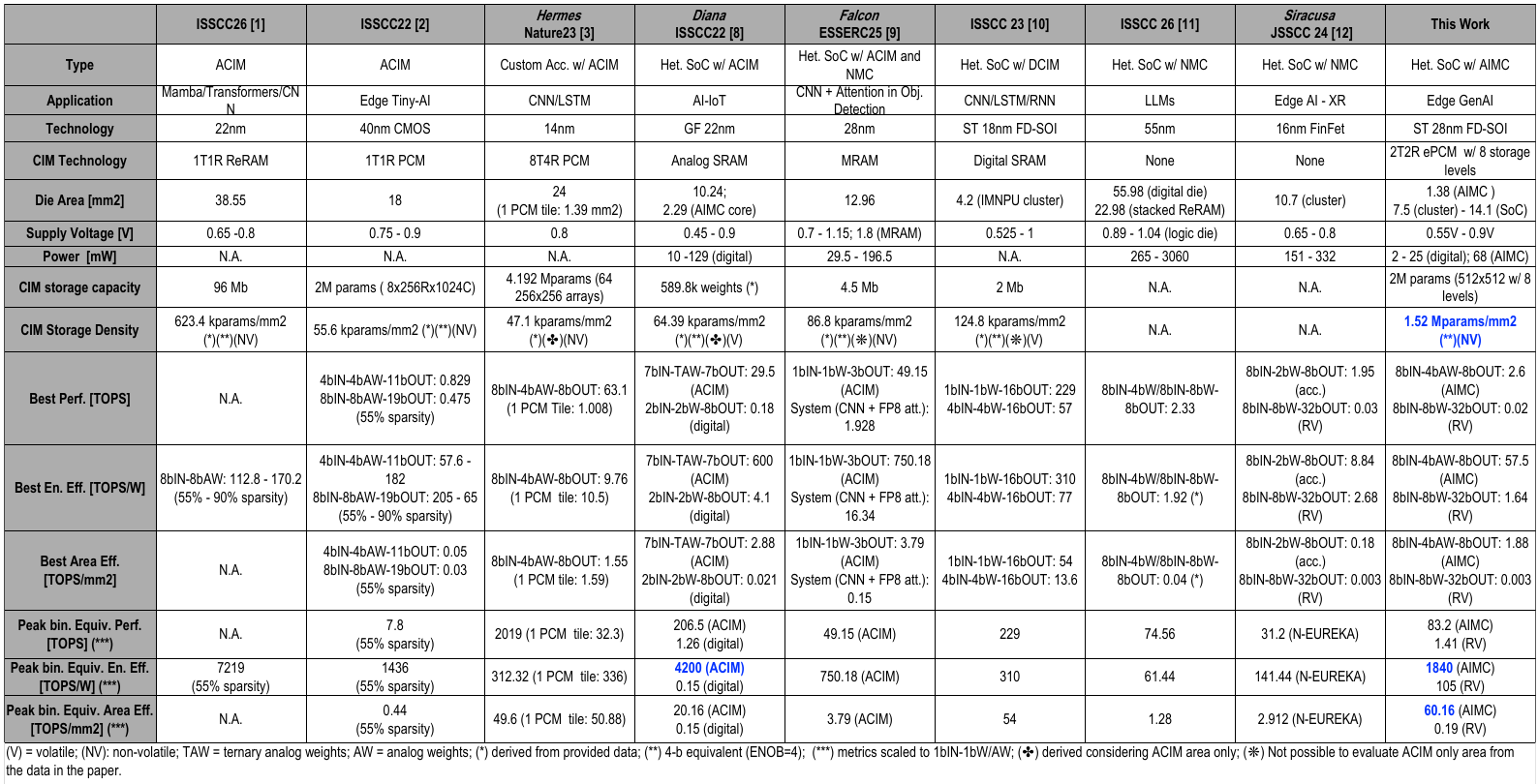}
    \caption{Comparison with state-of-the-art edge AI systems.}
    \label{fig:soa_comparison}
\end{figure*}

To contextualize the PCM/digital integration technique and \textsc{Megatron} capabilities, we evaluated \textit{SmolLM-135M} by HuggingFace as a representative Edge GenAI workload. Since PCM is non-volatile, the largest on-chip deployable network is capped by storage capacity (2Mparam for \textsc{Megatron}). We therefore modeled \textsc{Megatron-XL}, a larger configuration hosting two multiplexed PCiM macros to fit one SmolLM-135M layer ($3.54\times10^6$ parameters) and a larger 8MiB on-chip L2 SRAM for residuals and KV-cache. Data transfers were modeled both on-chip (L2--L1) and off-chip (L3--L1, for parameters without PCiM), using measured on-chip performance and power, and an LPDDR6 device~\cite{lpddr6} at 60~GB/s and 5~pJ/bit for off-chip access. Weights were 4-bit quantized with PCiM-characterized noise, and network accuracy was evaluated in PyTorch using \texttt{transformers} and \texttt{brevitas}, with analog noise modeled following Antolini~et~al.\cite{antolini}.

Figure~\ref{fig:end_to_end} reports accuracy, performance (time-to-first-token and time/token), and energy for SmolLM-135M. First, Fig.~\ref{fig:end_to_end}a confirms that quantization and analog noise do not significantly degrade accuracy on three representative benchmarks. In prefill (Fig.~\ref{fig:end_to_end}b left), PCiM yields $\sim60\times$ speedup on static-weight layers; combined with Softex ($\sim14\times$
 on non-linearities), this gives $\sim2.5\times$ overall per-layer speedup, with similar gains for what concerns energy; L3 access in the absence of PCM has limited impact due to on-chip data reuse in compute-bound conditions.
% 
% When we observe decode performance (Fig.~\ref{fig:end_to_end}b right), the pattern is similar.
% However, decode is typically memory-bound due to the number of matrix-vector operations.
% Consequently, the introduction of PCiM improves performance even more than in prefill phase: static weights on the non-volatile memory do not need to be loaded from L3, therefore the system is able to keep working in a compute-bound condition. Overall, \textsc{Megatron-XL} improves time/token by $\sim$\textbf{\color{red} $3.3\times$} compared to only the RISC-V cluster with no non-volatile memory.
% This is further detailed in Fig.~\ref{fig:end_to_end}c, where the decode latency of each operator in a SmolLM-135M layer is split in compute and DMA contributions.
% This effect is even more starking when considering energy, which in memory-bound conditions includes a very significant fraction due to L3 access.
% Overall, we achieve $\sim$\textbf{\color{red} $10.5\times$} better energy efficiency than the baseline RISC-V system.
In decode (Fig.~\ref{fig:end_to_end}b right), gains are even larger: decode is typically memory-bound, but PCiM eliminates L3 weight loading by keeping weights on-chip, restoring compute-bound operation. This yields $\sim3.3\times$ time/token improvement over the baseline RISC-V cluster, detailed per-operator in Fig.~\ref{fig:end_to_end}c. The energy benefit is particularly striking in memory-bound conditions due to eliminated L3 traffic, achieving $\sim10.5\times$ better energy efficiency overall.

\section{Comparison with State-of-the-art}

Fig.~\ref{fig:soa_comparison} compares \textsc{Megatron} against state-of-the-art edge-AI systems.
Considering non-volatile ReRAM-based~\cite{reram-transformer-isscc26} and PCM-based~\cite{pcm-isscc22} CiM-only SoC, we achieve 2.44$\times$ and 27.4$\times$ higher CiM storage density, respectively, thanks to our multi-level PCiM array, at comparable bit-equivalent energy efficiency, noting that their figures benefit from 55\% input sparsity, which we do not assume. Against \textit{Hermes}~\cite{hermes-nature}, which integrates 64 PCM-based ACiM tiles and a digital LSTM accelerator, we achieve comparable single-tile performance, while achieving overall comparable area efficiency, 5.5$\times$ higher energy efficiency, and 32.3$\times$ greater CiM storage density.
Compared to heterogeneous SoCs with SRAM-based~\cite{DIANA} and MRAM-based~\cite{falcon-esserc25} ACiM, we outperform both in area efficiency (2.98$\times$ and 15.9$\times$, respectively) and CiM storage density (23.6$\times$ and 17.5$\times$). While \textit{Diana}~\cite{DIANA} achieves higher peak energy efficiency, its SRAM-based ACiM lacks non-volatile parameter storage capabilities, potentially incurring higher end-to-end energy costs from memory-hierarchy parameter movement. Against a SRAM-based digital CiM SoC~\cite{desoli}, despite its higher throughput due to the integration of 64 32$\times$1024-b DCiM tiles, we achieve 1.2$\times$, 5.94$\times$ higher area and energy efficiency, respectively, with 12.2$\times$ greater storage density (non-volatile, in our case).
Finally, compared to traditional near-memory edge-ai computing SoCs~\cite{reram-on-logic-isscc26, siracusa}, we deliver 30$\times$ and 13$\times$ higher energy-efficiency, and 47$\times$ and 21$\times$ higher area-efficiency, respectively, at comparable or better performance, though their fully digital accelerators retain the advantage of iso-accuracy AI execution.

%% file: text/04_conclusion.tex
\section{Conclusion}
% We discussed the \textsc{Megatron} test chip, which demonstrates the feasibility of heterogeneous analog PCM/digital integration for Edge GenAI in a mature 28nm FD-SOI process.
% By combining a 2Mi-cell PCM compute-in-memory accelerator with a RISC-V DSP cluster and the Softex non-linear accelerator, the SoC achieves {57.5}~TOPS/W at {1.52}~Mparam/mm${}^2$ storage density.
% End-to-end benchmarking on a SmolLM-135M layer confirms that PCM-based in-memory computing substantially reduces both latency and energy, particularly in the memory-bound decode phase, with negligible accuracy degradation under analog noise.
% These results establish NV-AIMC as a viable path toward fully on-chip execution of Small Language Models at the edge.
The \textsc{Megatron} test chip demonstrates heterogeneous analog PCM/digital integration for Edge GenAI in 28nm FD-SOI, combining a 4Mi-cell, 2Mparam PCM compute-in-memory accelerator, RISC-V DSP cluster, and Softex non-linear accelerator to achieve {57.5}~TOPS/W at {1.52}~Mparam/mm${}^2$. SmolLM-135M benchmarking confirms PCM-based in-memory computing reduces latency and energy-especially in the memory-bound decode phase-with negligible accuracy loss, establishing NV-AIMC as a viable path to fully on-chip Small Language Model execution at the edge.